# Critical behavior of two-dimensional intrinsically ferromagnetic semiconductor CrI$_3$

G. T. Lin[1,2], X. Luo[1*], F. C. Chen[1,2], J. Yan[1,2], J. J. Gao[1,2], Y. Sun[3,2], W. Tong[3], P. Tong[1], W. J. Lu[1], Z. G. Sheng[3,4], W. H. Song[1], X. B. Zhu[1], and Y. P. Sun[3,1,4*]

[1] *Key Laboratory of Materials Physics, Institute of Solid State Physics, Chinese Academy of Sciences, Hefei, 230031, China*

[2] *University of Science and Technology of China, Hefei, 230026, China*

[3] *High Magnetic Field Laboratory, Chinese Academy of Sciences, Hefei, 230031, China*

[4] *Collaborative Innovation Center of Advanced Microstructures, Nanjing University, Nanjing, 210093, China*

[*]Corresponding authors: xluo@issp.ac.cn and ypsun@issp.ac.cn.

**Abstract**

$CrI_3$, which belongs to a rare category of two-dimensional (2D) ferromagnetic semiconductors, is of great interest for spintronic device applications. Unlike $CrCl_3$ whose magnetism presents a 2D-Heisenberg behavior, $CrI_3$ exhibits a larger van der Waals gap, smaller cleavage energy, and stronger magnetic anisotropy which could lead to a 3D magnetic characteristic. Hence, we investigate the critical behavior of $CrI_3$ in the vicinity of magnetic transition. We use the modified Arrott plot and Kouvel-Fisher method, and conduct critical isotherm analysis to estimate the critical exponents near the ferromagnetic phase transition. This shows that the magnetism of $CrI_3$ follows the crossover behavior of a 3D-Ising model with mean field type interactions where the critical exponents $\beta$, $\gamma$, and $\delta$ are $0.323 \pm 0.006$, $0.835 \pm 0.005$, and $3.585 \pm 0.006$, respectively, at the Curie temperature of 64 K. We propose the crossover behavior can be attributed to the strong uniaxial anisotropy and inevitable interlayer coupling. Our experiment demonstrates the applicability of crossover behavior to a 2D ferromagnetic semiconductor.

Since the discovery of single layer graphene, two-dimensional (2D) materials have displayed a broad range of electronic properties and immense potential in spintronic applications due to their highly tunable physical properties [1-7]. However, spintronic devices using 2D materials are still in their preliminary stage [3,8-11], which is due to the lack of long-range ferromagnetic order that is crucial for spintronic applications [5-7]. The appearance of ferromagnetism in 2D materials as well as their rich electrical and optical properties creates ample opportunities for fundamental science discoveries along with the possible seamless integration of information processing and storage [5-7].

Recently, Chromium trihalides Cr$X_3$ ($X$ = Cl, Br, and I) with the rhombohedral BiI$_3$ structure at low temperatures have been widely studied because they belong to a rare category of ferromagnetic semiconductors with a 2D layered structure [6-9,12-21]. On the theoretical level, recent studies on Cr$X_3$ focus on their electronic structure and magnetic properties, particularly predictions of the single-layer properties [9,17-21]. Meanwhile, the theory predicts that CrBr$_3$ has 3D-Ising magnetic characteristics [17,20]. On the experimental level, CrCl$_3$ shows characteristics of 2D-Heisenberg behavior and antiferromagnetic order between layers [13,16]. With the increase of the $X$ atom radius, Cr$X_3$ presents a larger van der Waals gap and in-plane nearest-neighbor Cr-Cr distance, which enhances the Curie temperature from about 17 K for CrCl$_3$, 37 K for CrBr$_3$ to 61 K for CrI$_3$ [6,8,13-16]. In addition, a scanning magneto-optic Kerr microscopy experiment shows that monolayer CrI$_3$ is an Ising ferromagnet with out-of-plane spin orientation. With thickness increases, there appears to be a layer-dependent magnetic phase [6].

It is known that, with the increase of the $X$ atom radius, Cr$X_3$ shows larger van der Waals gap and smaller cleavage energy [18,19]. We suppose that the Cr$X_3$ system may undergo a magnetic phase transition from a 2D-Heisenberg ferromagnet to 3D-Ising behavior with the increase of the $X$ atom radius. Therefore, a method to rapidly characterize the critical behavior of single-crystalline CrI$_3$ is crucial. For this purpose, we present a detailed investigation of the critical phenomena of CrI$_3$ using the initial isothermal $M(H)$ curves around the Curie temperature $T_C$. We find that the strong magnetic anisotropy and inevitable interlayer coupling can induce CrI$_3$ to show 3D magnetic characteristics where the crossover behavior is consistent with a 3D-Ising model with long range or mean field type interactions.

Single-crystalline CrI$_3$ was prepared by the chemical vapor transport method [8]. The single-crystalline x-ray diffraction (XRD) data collected at room temperature indicated that CrI$_3$ was monoclinic structure, which is consistent with Ref. [8] (see the Supplemental Material). We measured the heat capacity using the Quantum Design physical properties measurement system (PPMS-9T) and characterized the magnetic properties by the magnetic property measurement system (MPMS-XL5).

Figure 1(a) and (b) show the temperature-dependent inverse susceptibility $1/\chi(T)$ of CrI$_3$ under zero-field cooled warming with applied magnetic field $H = 100$ Oe, parallel to the $c$ axis and $ab$ plane, respectively. We observe a paramagnetic–ferromagnetic (PM-FM) transition at the $T_\text{C}^\text{mag}$ of 61 K, as determined by the derivative of the susceptibility, which is consistent with the values of 61-68 K reported previously [8,12]. Meanwhile, the anomaly, observed in the paramagnetic area, reveals a first-order crystallographic phase transition occurring near the $T_S$ of 211 K [8]. For a FM system, the $\chi(T)$ above $T_C^{mag}$ can be described by the Curie-Weiss law resulting from the mean-field theory [22]. The $1/\chi(T)$ data are near-linear behavior over a relatively wide range of temperatures both above and below $T_S$, which is fitted by the Curie-Weiss law (shown in Figure 1(a) and (b)). The effective magnetic moment $\mu_{eff}$ determined from the Curie constants (C) is about $4\mu_B$ both above and below $T_S$, which is close to the theoretical value expected for $Cr^{3+}$ of 3.87 $\mu_B$ [10]. The insets of Figure 1(a) and (b) show the isothermal magnetization $M(H)$ at 5 K exhibiting a typical FM behavior with the saturation field $H_S$ of about 2.3 kOe (parallel to the $c$ axis) and 31 kOe (parallel to the $ab$ plane), which confirm that there is an obvious magnetic anisotropy in CrI$_3$. The saturation moment is about 3 $\mu_B$/Cr, close to the theoretical value expected for high spin $Cr^{3+}$ (S = 3/2). In addition, the $M(H)$ curves show almost no coercive force for CrI$_3$, implying it has obvious soft FM features.

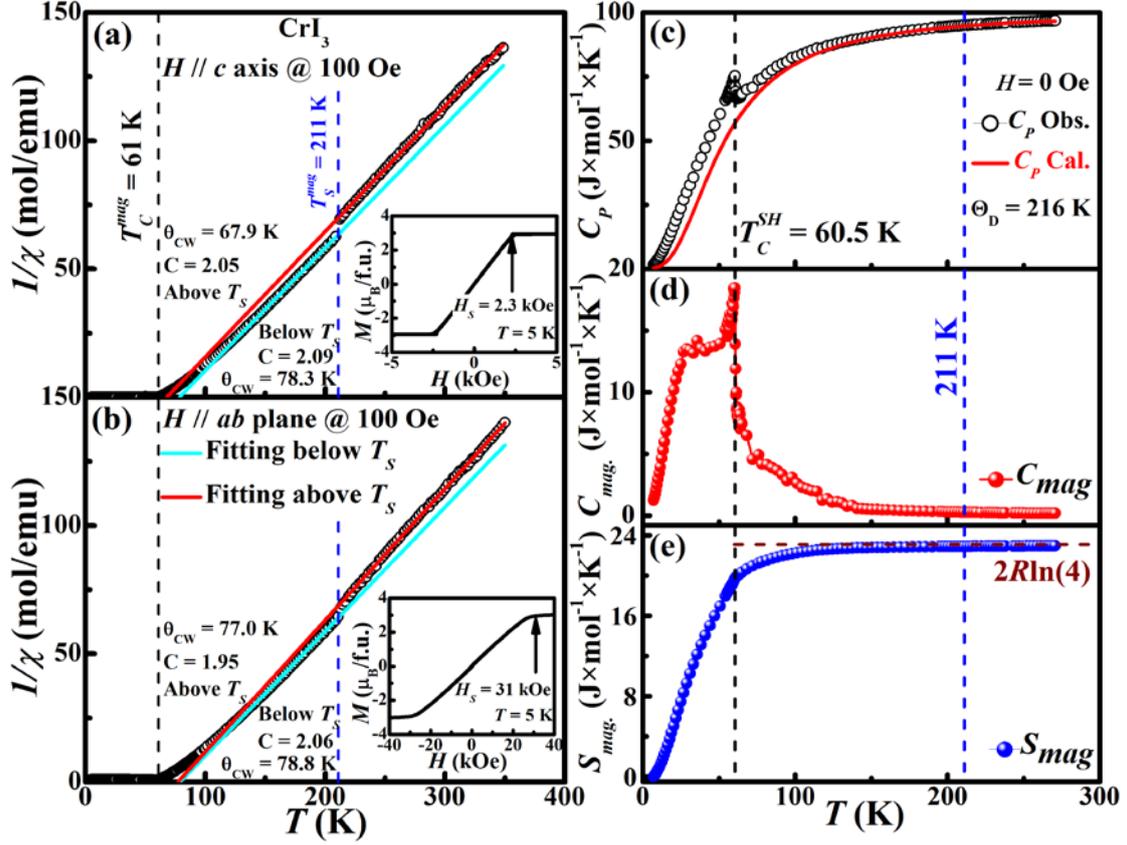

**Figure 1.** (Color online): (a) and (b) Temperature-dependent inverse susceptibility $1/\chi(T)$ of CrI$_3$ under zero field cooled warming at 100 Oe, parallel to the *c* axis and *ab* plane, respectively. The red and cyan solid lines are the fitted results according to the Curie-Weiss law. The insets show the isothermal magnetization curves *M(H)* at 5 K. (c) Specific heat $C_p$ as a function of *T* for CrI$_3$ and the fitted $C_V^{Debye}(T)$ using Eq. (1); Temperature-dependent magnetic (d) specific heat $C_{mag}(T)$ and (e) entropy $S_{mag}(T)$. The wine dashed line refers to $S_{mag}(T\to\infty)$ calculated with the magnetic moment S of Cr$^{3+}$ being 3/2.

Figure 1(c) shows the variation of the zero-field specific heat (SH) $C_p(T)$ under cooling-temperature mode. The sharp anomaly in $C_p(T)$ at 60.5 K corresponds to FM transition temperature $T_C^{SH}$. Since CrI$_3$ is a semiconductor, the electronic contribution to the heat capacity is not considered. The $C_{mag}$ can be calculated by the following equations :[22]

$$C_{mag}(T) = C_p(T) - N \times C_V^{Debye}(T), \qquad (1)$$

$$C_V^{Debye}(T) = 9 \times R \times \left(\frac{T}{\Theta_D}\right)^3 \int_0^{\Theta_D/T} \frac{x^4 \times e^x}{(e^x - 1)^2} dx, \qquad (2)$$

where $R$ is the molar gas constant, $\Theta_D$ is the Debye temperature, and $N$ = 4 is the number of atoms per formula unit. The sum of Debye functions accounts for the lattice contribution to SH. We can extract the magnetic contribution $C_{mag}(T)$ from the measured SH of CrI$_3$. The fitted $C_p(T)$ for CrI$_3$ by Eq. (1) and (2) over the temperature range from about 7 to 270 K is shown by the red curve in Figure 1(c) using the Debye temperature $\Theta_D$ = 216 K. The $C_{mag}(T)$ curve exhibits a sharp peak at $T_C^{SH}$ of 61 K. The magnetic entropy $S_{mag}(T)$ is calculated by $S_{mag}(T) = \int_0^T \frac{C_{mag}(T)}{T} dT$. Figure 1(e) shows the temperature dependence of $S_{mag}(T)$. The entropy of CrI$_3$ per mole with completely disordered spins S is $S_{mag}(T \to \infty) = 2 \times R \times \ln(2 \times S + 1)$. Using S = 3/2 for Cr$^{3+}$, we obtain $S_{mag}(T \to \infty)$ of 23.1 J/(mol×K).

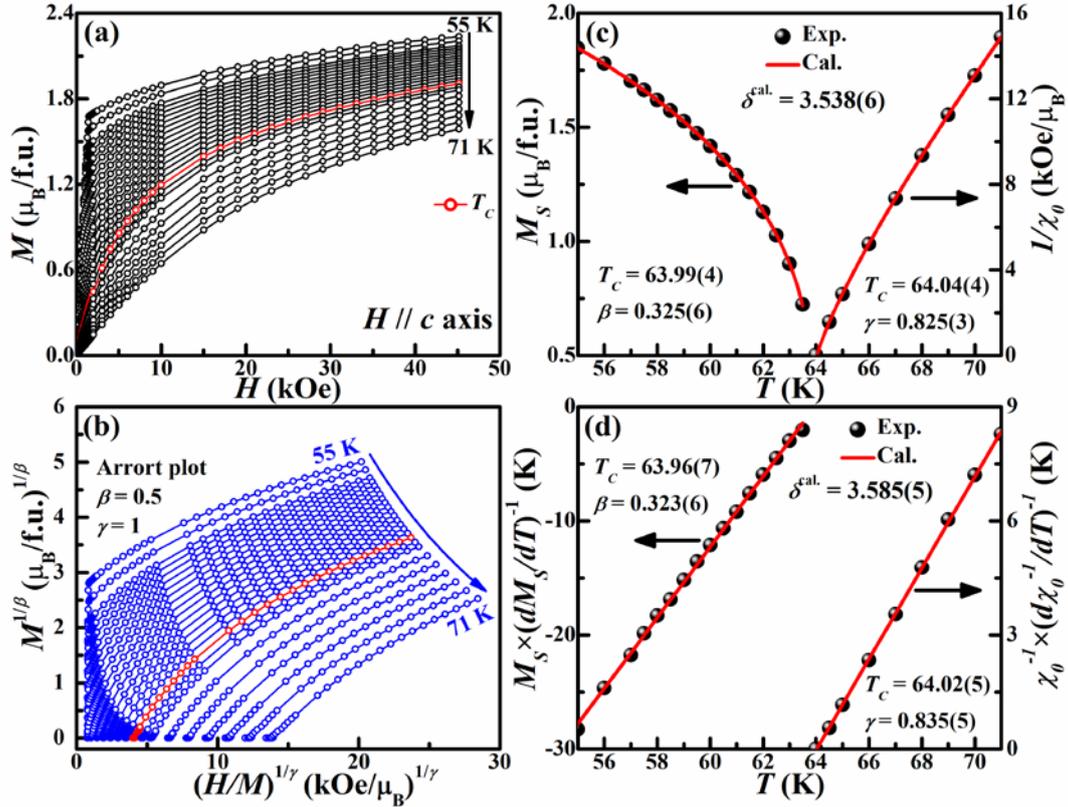

**Figure 2.** (Color online): (a) Isothermal magnetization data for selected temperatures around $T_C$; (b) Arrott plots of $M^2$ versus $H/M$; (c) Temperature dependence of $M_S$ and $\chi_0^{-1}$. The $T_C$ and critical exponents are obtained from the fitting of Eq. (S1) and (S2); (d) The Kouvel-Fisher plot. The $T_C$ and critical exponents are obtained from the linear fit.

As mentioned above, with the $X$ atom radius increases, the CrI$_3$ compounds present the larger van der Waals gap and the smaller cleavage energy, which may induce a 2D magnetic phase transition. For the purpose of confirmation, we performed a detailed characterization of the critical phenomena using the initial isothermal $M(H)$ curves around $T_C$ for the CrI$_3$, which is shown in Figure 2(a).

Figure 2(b) shows the Arrott plots for the compound $CrI_3$, where critical exponents follow mean-field behaviour with $\beta = 0.5$ and $\gamma = 1$. And the positive slope for full temperature and field range of investigation reveals a second-order phase transition according to the criterion set by Banerjee [23]. For mean-field type critical behavior, it is expected that the high field magnetization should be a series of parallel straight lines around $T_C$. Further, the critical isotherm at $T = T_C$ should pass through the origin. However, all the curves in Figure 2(b) show nonlinear behaviors, which indicates the mean field model with $\beta = 0.5$ and $\gamma = 1$ is not suitable exponents for this system. As such, a modified Arrott plot (MAP) should be employed to obtain the critical exponents.

In order to extract the accurate critical exponents for this system, we usually employ the MAP technique based on Arrott-Noaks equation of state following Eq. (S5) for $CrI_3$ at different temperatures (see the supporting information). By proper selections of $\beta$ and $\gamma$, one can clearly show the isotherms are a set of parallel straight lines in high fields as displayed in the Figure S2. The linear extrapolation from the high field region gives the spontaneous magnetization $M_S(T,0)$ and the inverse of initial susceptibility $\chi_0^{-1}(T,0)$ (see Figure 2(c)) corresponding the intercepts on the $M^{1/\beta}$ and $(H/M)^{1/\gamma}$ axes, respectively. By fitting the data of $M_S(T,0)$ and $\chi_0^{-1}(T,0)$ to Eq. (S1) and (S2), we obtain two new values of $\beta = 0.325 \pm 0.006$ with $T_C = 63.99 \pm 0.04$ and $\gamma = 0.825 \pm 0.003$ with $T_C = 64.04 \pm 0.04$. The estimated $\beta$ value is close to 3D-Ising behavior whereas $\gamma$ value signifies long range ordering with mean field type behavior. In addition, these critical exponents and $T_C$ can be deduced more accurately from the Kouvel-Fisher (KF) method [24]. Hence, one can find that the temperature dependence of $M_S \times (dM_S/dT)^{-1}$ and $\chi_0^{-1} \times (d\chi_0^{-1}/dT)^{-1}$ should be straight lines with slopes $1/\beta$ and $1/\gamma$, respectively. As seen in Figure 2(d), the linear fit yields the $\beta$ of $0.323 \pm 0.006$ with $T_C$ of $63.96 \pm 0.07$ and $\gamma$ of $0.835 \pm 0.005$ with $T_C$ of $64.02 \pm 0.05$, respectively, which agrees with estimation from MAP technique.

Moreover, the third exponent $\delta$ can be calculated by the Widom scaling relation [25,26]

$$\delta = 1 + \frac{\gamma}{\beta}. \tag{3}$$

Based on the $\beta$ and $\gamma$ values calculated in Figure 2(c) and (d), Eq. (3) yields $\delta$ of $3.538 \pm 0.006$ and $3.585 \pm 0.006$, respectively. Figure 3(a) shows the isothermal magnetization $M(H)$ at $T_C = 64$ K. According to Eq. (S3), $\delta = 3.569 \pm 0.004$ can be directly estimated from the critical isotherm (CI) at $T_C$, which is very close to the values obtained from the modified Arrott plot and KF method. The inset of the same plot has been demonstrated on a log-log scale. The solid straight line with a slope $1/\delta$ is the fitted result using Eq. (S3). This suggests that the estimated values are self-consistent and unambiguous.

Finally, the reliability of the obtained critical exponents should follow the scaling equation (Eq. (S6)) in the critical region. The scaling equation indicates that $m$ versus $h$ forms two universal curves for $T > T_C$ and $T < T_C$, respectively. Based on Eq. (S7), the isothermal magnetization around the critical temperatures for $CrI_3$ has been plotted in Figure 3(b). It is quite significant that all experimental data in the higher-field region collapse onto two separate branches. The inset of Figure 3(b) shows the corresponding log-log plot. Similarly, all the points also collapse into two different curves in the higher-field region. Meanwhile, we examine the convergence of the critical exponents, the effective exponents $\beta_{eff}$ and $\gamma_{eff}$ can be obtained by Eq. (S8) and (S9) for CrI$_3$. As shown in Figure 3(c) and (d), both $\beta_{eff}$ and $\gamma_{eff}$ show a non-monotonic variation with $\varepsilon$ (see Eq. (S4)). The lowest $\varepsilon$ ($\varepsilon_{min}$) are $7.19\times10^{-3}$ and $7.18\times10^{-3}$ for $\beta_{eff}$ and $\gamma_{eff}$, respectively. We obtain the effective exponents $\beta_{eff}$ of 0.325 and $\gamma_{eff}$ of 0.833, indicating that both $\beta_{eff}$ and $\gamma_{eff}$ converge when the temperature approaches $T_C$. The well-rescaled curves further confirm that the obtained results of the critical exponents and $T_C$ are valid.

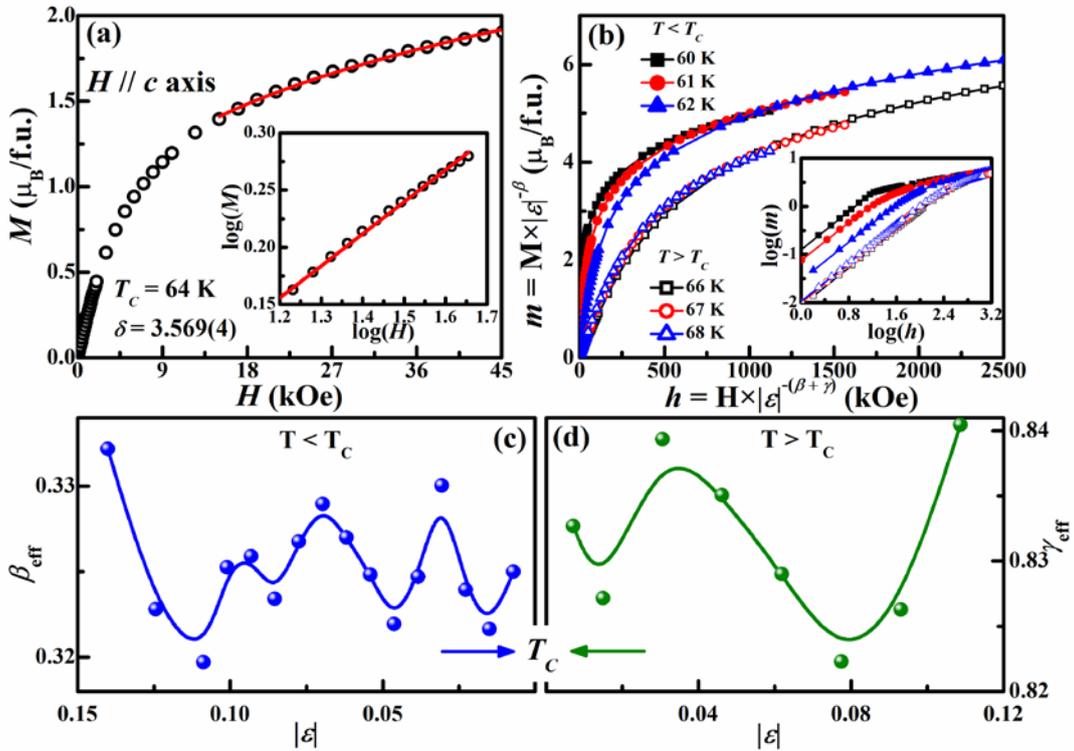

**Figure 3.** (Color online): (a) Isothermal $M(H)$ at $T_C$. The inset shows the alternative plot on a log-log scale and the straight line is the linear fit following Eq. (S3); (b) Renormalized magnetization $m$ versus renormalized field $h$ at several typical temperatures around the $T_C$. The inset shows the alternative plot on a log-log scale; the effective exponents (c) $\beta_{eff}$ below $T_C$ and (d) $\gamma_{eff}$ above $T_C$ as a function of the reduced temperature $\varepsilon$.

In order to further confirm the magnetic behavior presents the 3D-Ising model in CrI$_3$, the SH curves in the near vicinity of the Néel temperature have been carefully measured with high resolution. The experimental SH data have been fitted to the well-known relation [27], which is shown in Figure 4:

$$C_P = B + C \times \varepsilon + A^{\pm} \times |\varepsilon|^{-\alpha} \times \left(1 + E^{\pm} \times |\varepsilon|^{0.5}\right), \quad (4)$$

where $\alpha$, $A^{\pm}$, $B$, $C$, and $E^{\pm}$ are adjustable parameters. Superscripts + and − stand for $T > T_C$ and $T < T_C$, respectively. The relevant information about fitting procedures can be found in the following reference [27]. By Eq. (4), we obtain the value of critical exponent $\alpha = 0.11$ and the ratio of the critical coefficients $A^+/A^- = 0.49(4)$, which is close to the 3D-Ising universality class. Hence, the observed values from SH are in concert with our estimation from magnetization measurements.

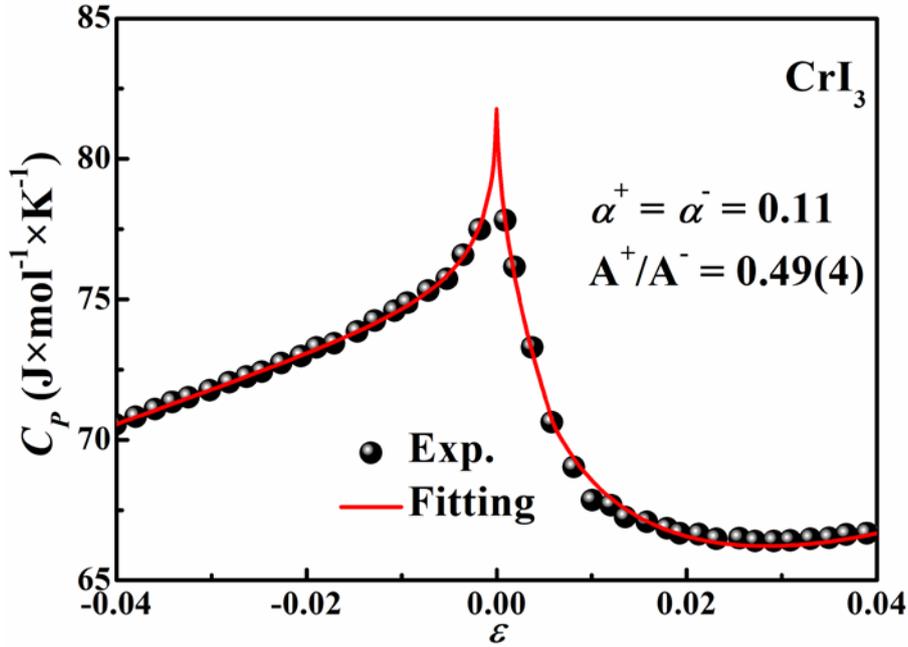

**Figure 4.** (Color online): Experimental (black circles) and fitted curves (red line) of the SH as a function of the reduced temperature in the vicinity of $T_C$ by using Eq. (4).

All the critical exponents estimated from various techniques as well as the known theoretical model values are summarized in Table 1. Taroni et al. have accomplished a comprehensive investigation of critical exponents that the critical exponent $β$ should be within a window ∼$0.1 ≤ β ≤ 0.25$ for 2D magnets [28]. That is to say, the magnetic interactions of CrI$_3$ exhibit clear 3D characteristic, which indicates that the interlayer coupling should be considerable. As we mentioned above, the $β$ value indicates that CrI$_3$ possesses 3D-Ising model whereas $γ$ value is close to mean-field model and $δ$ is located between 3D-Ising and mean-field model. The mean-field-like values suggest that long-range coupling should be important in CrI$_3$ [29]. These exponents close to the 3D-Ising model can be attributed to uniaxial anisotropy causing a crossover effect, which is completely consistent with spontaneous magnetization along the $c$ axis [6,8,9]. Crossover behavior has become a ubiquitous phenomenon, such as the doped La$_{0.75}$Sr$_{0.25}$MnO$_3$ sample in which the critical exponents are between 3D-Ising model values and mean-field ones [29]. The crossover behavior shows a 3D-Ising model values coupled with long-range spin interactions in Fe$_3$GeTe$_2$ [32]. Recently, polycrystalline $Gd_2Cu_2In$ has also been reported that the $β$ value is close to the 3D-Ising model and $γ$ and $δ$ values are close to mean-field model [33]. Hence, our results confirm a possible direct crossover from the mean-field model to the 3D-Ising model in the vicinity of the phase transition for $CrI_3$.

**Table 1.** Critical exponents of CrI$_3$ with various theoretical models (SC = single crystal; cal = Calculated from Eq. (3)).

| Composition | Technique | $β$ | $γ$ | $δ$ | $α$ | $A^+/A^-$ | Ref. |
|---|---|---|---|---|---|---|---|
| CrI$_3^{SC}$ | MAP | 0.325 ± 0.006 | 0.825 ± 0.003 | 3.538 ± 0.006$^{cal}$ | - | - | This work |
|  | KF | 0.323 ± 0.006 | 0.835 ± 0.005 | 3.585 ± 0.006$^{cal}$ | - | - |  |
|  | CI | - | - | 3.569 ± 0.004 | - | - |  |
|  | SH | - | - | - | 0.11 | 0.49(4) |  |
| Tricritical mean-field | Theory | 0.25 | 1 | 5 | - | - | [30,31] |
| Mean-field | Theory | 0.5 | 1 | 3 | 0 | - |  |
| 3D-Ising | Theory | 0.325 | 1.24 | 4.82 | 0.11 | 0.52 |  |

In addition, the critical behavior of 2D magnetic materials has been widely studied. For example, compared to $CrSiTe_3$ with 2D-Ising ferromagnetic behavior [34], $CrGeTe_3$ exhibits tricritical mean-field model [10]. Obviously，though both Cr*M*$Te_3$ (*M* = Si, Ge, Sn) and Cr*X*$_3$ crystallize as a rhombohedral structure with the space group $R\bar{3}$ at low temperature [35-37], their magnetism shows different critical behaviors [10,34], which is closely related to their interlayer coupling. Comparing the interlayer distance between Cr*M*$Te_3$ and Cr*X*$_3$, one can observe that Cr*M*$Te_3$ system (3.32 Å for $CrSiTe_3$ and 3.27 Å for $CrGeTe_3$) has a larger van der Waals gap than that of Cr*X*$_3$ (2.698 Å for $CrCl_3$, 2.909 Å for $CrBr_3$, and 3.174 Å for $CrI_3$) [18,35]. Therefore, we can deduce that the larger interlayer interaction may be one of the reasons that despite of 2D structural characteristics, there are distinct 3D magnetic properties in the Cr*X*$_3$ system. For $CrCl_3$, the magnetic ground state presents that the magnetic moments are aligned ferromagnetically within layers and the layers are stacked antiferromagnetically [13], which indicates strong spin interactions between layers. With the increase of the *X* atom radius, Cr*X*$_3$ presents the larger van der Waals gap, the smaller cleavage energy [18,19] and stronger magnetic anisotropy [6,8,9,13,38], which lead to the magnetic ground state transition from antiferromagnetic order between layers to ferromagnetic order along the *c* axis [6,8,9,13,38]. Hence, the strong magnetic anisotropy and inevitable interlayer coupling could induce a crossover behavior with the 3D-Ising ferromagnetic characteristics in the single-crystalline $CrI_3$.

In conclusion, we have presented comprehensive and detailed studies on the critical properties of single-crystalline $CrI_3$ using isothermal magnetization in the vicinity of its magnetic transition. Based on various experimental techniques including the MAP, KF method and CI analysis, we obtain the critical exponents $β$, $γ$, and $δ$ of $0.323 ± 0.006$, $0.835 ± 0.005$, and $3.585 ± 0.006$, respectively, at the Curie temperature of 64 K. The spontaneous magnetization exponent $β$ value suggests that the system possesses 3D-Ising behavior whereas $γ$ and $δ$ values indicate long range or mean field type interactions. The crossover behavior can be attributed to the strong uniaxial anisotropy and inevitable interlayer coupling, which is completely consistent with spontaneous magnetization along the $c$ axis.

**Supplementary Material**

See supplementary material for the crystal structure information and the complete scaling analysis of critical behavior in the vicinity of magnetic transition.

**Acknowledgements**

This work was supported by the National Key Research and Development Program under contract 2016YFA0300404 and 2016YFA0401803, and the Joint Funds of the National Natural Science Foundation of China and the Chinese Academy of Sciences' Large-Scale Scientific Facility under contract U1432139, the National Nature Science Foundation of China under contract 11674326, and Key Research Program of Frontier Sciences, CAS (QYZDB-SSW-SLH015), and the Nature Science Foundation of Anhui Province under contract 1508085ME103 and Hefei Science Center of CAS under contract 2016HSC-IU011. The first author thanks Dr. C. Sun from University of Wisconsin-Madison for her assistance in editing the revised manuscript.

**References**

[1] K. S. Novoselov, A. K. Geim, S. V. Morozov, D. Jiang, Y. Zhang, S. V. Dubonos, I. V. Grigorieva, and A. A. Firsov, science **306**, 666 (2004).

[2] K. S. Novoselov, A. K. Geim, S. V. Morozov, D. Jiang, M. I. Katsnelson, I. V. Grigorieva, S. V. Dubonos, and A. A. Firsov, Nature **438**, 197 (2005).

[3] W. Han, R. K. Kawakami, M. Gmitra, and J. Fabian, Nat. Nanotechnol. **9**, 794 (2014).

[4] X. Xu, W. Yao, D. Xiao, and T. F. Heinz, Nat. Phys. **10**, 343 (2014).

[5] C. Gong, L. Li, Z. Li, H. Ji, A. Stern, Y. Xia, T. Cao, W. Bao, C. Wang, Y. Wang *et al.*, Nature **546**, 265 (2017).

[6] B. Huang, G. Clark, E. Navarro-Moratalla, D. R. Klein, R. Cheng, K. L. Seyler, D. Zhong, E. Schmidgall, M. A. McGuire, D. H. Cobden *et al.*, Nature **546**, 270 (2017).

[7] D. Zhong, K. L. Seyler, X. Linpeng, R. Cheng, N. Sivadas, B. Huang, E. Schmidgall, T. Taniguchi, K. Watanabe, M. A. McGuire *et al.*, Science Adv. **3** (2017).

[8] M. A. McGuire, H. Dixit, V. R. Cooper, and B. C. Sales, Chem. Mater. **27**, 612 (2015).

[9] J. L. Lado and J. Fernández-Rossier, 2D Mater. **4**, 035002 (2017).

[10] G. T. Lin, H. L. Zhuang, X. Luo, B. J. Liu, F. C. Chen, J. Yan, Y. Sun, J. Zhou, W. J. Lu, P. Tong *et al.*, Phys. Rev. B **95** (2017).

[11] Y. Liu and C. Petrovic, Phys. Rev. B **96** (2017).

[12] J. F. Dillon and C. E. Olson, J. Appl. Phys. **36**, 1259 (1965).

[13] M. A. McGuire, G. Clark, S. Kc, W. M. Chance, G. E. Jellison, V. R. Cooper, X. Xu, and B. C. Sales, Physical Review Materials **1** (2017).

[14] E. J. Samuelsen, R. Silberglitt, G. Shirane, and J. P. Remeika, Phys. Rev. B **3**, 157 (1971).

[15] C. H. Cobb, V. Jaccarino, J. P. Remeika, R. Silberglitt, and H. Yasuoka, Phys. Rev. B **3**, 1677 (1971).

[16] M. Suzuki and I. S. Suzuki, Phys. Rev. B **57**, 10674 (1998).

[17] D. J. Klein and N. H. March, Phys. Lett. A **372**, 5052 (2008).

[18] J. Liu, Q. Sun, Y. Kawazoe, and P. Jena, Phys. Chem. Chem. Phys. **18**, 8777 (2016).

[19] W.-B. Zhang, Q. Qu, P. Zhu, and C.-H. Lam, Journal of Materials Chemistry C **3**, 12457 (2015).

[20] Z. D. Zhang, Philos. Mag. **87**, 5309 (2007).

[21] H. L. Davis and A. Narath, Phys. Rev. **134**, A433 (1964).

[22] C. Kittel, *Introduction to Solid State Physics* (2008), New York: Wiley.

[23] B. Banerjee, Physics letters **12**, 16 (1964).

[24] J. S. Kouvel and M. E. Fisher, Phys. Rev. **136**, A1626 (1964).

[25] B. Widom, The Journal of Chemical Physics **41**, 1633 (1964).

[26] B. Widom, The Journal of Chemical Physics **43**, 3898 (1965).

[27] A. Oleaga, A. Salazar, D. Prabhakaran, J. G. Cheng, and J. S. Zhou, Phys. Rev. B **85** (2012).

[28] A. Taroni, S. T. Bramwell, and P. C. Holdsworth, Journal of physics. Condensed matter : an Institute of Physics journal **20**, 275233 (2008).

[29] D. Kim, B. L. Zink, F. Hellman, and J. M. D. Coey, Phys. Rev. B **65** (2002).
[30] S. Kaul, J. Magn. Magn. Mater. **53**, 5 (1985).
[31] A. Arrott, Phys. Rev. **108**, 1394 (1957).
[32] B. Liu, Y. Zou, S. Zhou, L. Zhang, Z. Wang, H. Li, Z. Qu, and Y. Zhang, Sci. Rep. **7**, 6184 (2017).
[33] H. S. N. K. Ramesh Kumar, B. N. Sahu, Sindisiwe Xhakaza, and Andre M. Strydom, arXiv:1711.09816 (2017).
[34] B. Liu, Y. Zou, L. Zhang, S. Zhou, Z. Wang, W. Wang, Z. Qu, and Y. Zhang, Sci. Rep. **6**, 33873 (2016).
[35] X. Li and J. Yang, Journal of Materials Chemistry C **2**, 7071 (2014).
[36] V. Carteaux, F. Moussa, and M. Spiesser, EPL (Europhysics Letters) **29**, 251 (1995).
[37] V. Carteaux, D. Brunet, G. Ouvrard, and G. Andre, J. Phys.: Condens. Matter **7**, 69 (1995).
[38] J. J. F. Dillon, J. Phys. Soc. Jpn. **19**, 1662 (1964).